\input harvmac
\input epsf
\def\figin{\epsfcheck\figin}\def\figins{\epsfcheck\figins}
\def\epsfcheck{\ifx\epsfbox\UnDeFiNeD
\message{(NO epsf.tex, FIGURES WILL BE IGNORED)}
\gdef\figin##1{\vskip2in}\gdef\figins##1{\hskip.5in}
\else\message{(FIGURES WILL BE INCLUDED)}%
\gdef\figin##1{##1}\gdef\figins##1{##1}\fi}
\def\DefWarn#1{}
\def\figinsert{\goodbreak\midinsert}
\def\ifig#1#2#3{\DefWarn#1\xdef#1{fig.~\the\figno}
\writedef{#1\leftbracket fig.\noexpand~\the\figno}%
\figinsert\figin{\centerline{#3}}\medskip\centerline{\vbox{\baselineskip12pt
\advance\hsize by -1truein\noindent\footnotefont{\bf Fig.~\the\figno:} #2}}
\bigskip\endinsert\global\advance\figno by1}

\lref\geomeng{S. Katz, A. Klemm, C. Vafa, {\it Geometric Engineering of Quantum Field Theories},  EFI-96-37, HUTP-96/A046, OSU-M-96-24, hep-th/9609239.}

\lref\lerchewarner{W. Lerche, N. P. Warner, {\it Exceptional SW Geometry from ALE Fibrations}, CERN-TH/96-237, USC-96/022, hep-th/9608183.}

\lref\kv{S. Kachru, C. Vafa, {\it Exact Results for N=2 Compactifications of Heterotic Strings}, Nucl. Phys. B450 (1995) 69, hep-th/9505105; S. Kachru, A. Klemm, W. Lerche, P. Mayr, C. Vafa, {\it Nonperturbative Results on the Point Particle 
Limit of N=2 Heterotic String Compactifications},  Nucl.Phys. B459 (1996) 537, 
hep-th/9508155.}

\lref\aspinwallgross{ P. S. Aspinwall, M. Gross, {\it 
The SO(32) Heterotic String on a K3 Surface}, Phys.Lett. B387 (1996) 735, 
 hep-th/9605131.}

\lref\bershadskyetal{M. Bershadsky, K. Intriligator, S. Kachru, 
D. R. Morrison, V. Sadov, C. Vafa, {\it  
Geometric Singularities and Enhanced Gauge Symmetries}, Nucl.Phys.B481:215-252,1996, hep-th/9605200.} 

\lref\selfdualstrings{ A. Klemm, W. Lerche, P. Mayr, C. Vafa, N. Warner, 
{\it Self-Dual Strings and N=2 Supersymmetric Field Theory}, 
Nucl.Phys. B477 (1996) 746-766, hep-th/9604034.}

\lref\martinecwarner{E. Martinec, N. Warner, {\it Integrable systems and supersymmetric gauge theory}, Nucl.Phys. B459 (1996) 97, hep-th/9509161.}

\lref\giddings{S. B. Giddings, J. M. Pierre, {\it Some exact results in supersymmetric theories based on
exceptional groups},  Phys.Rev. D52 (1995) 6065-6073, hep-th/9506196.}

\lref\cvitanovic{Cvitanovic, {\it Group Theory: Classics Illustrated}.}

\lref\toric{P.S. Aspinwall, B.R. Greene and D.R. Morrison, {\it 
Calabi-Yau Moduli Space, Mirror Manifolds and Spacetime Topology Change in
String Theory},  IASSNS-HEP-93/38, CNLS-93/1236, hep-th/9309097; 
S. Hosono, A. Klemm, S. Theisen, 
{\it Lectures on Mirror Symmetry}, LMU-TPW-94-02,hep-th/9403096; 
S. Hosono (Harvard U.), A. Klemm, S. Theisen (Munich U., Theor. Phys.), 
S.T. Yau, {\it 
Mirror Symmetry, Mirror Map and Applications to Calabi-Yau Hypersurfaces}
Commun. Math. Phys. 167 (1995) 301, hep-th/9308122.}

\lref\lerchereview{W. Lerche, {\it Introduction to Seiberg-Witten Theory and its Stringy Origin}, CERN-TH/96-332, hep-th/9611190.}

\lref\mv{D. R. Morrison, C. Vafa, {\it Compactifications of F-Theory on 
Calabi--Yau Threefolds
-- I},  Nucl.Phys. B473 (1996) 74-92, hep-th/9602114; {\it 
Compactifications of F-Theory on Calabi--Yau Threefolds
-- II}, 
Nucl.Phys. B476 (1996) 437, hep-th/9603161}

\lref\f{C. Vafa, {\it Evidence for F-Theory}, Nucl.Phys. B469 (1996) 
403-418, hep-th/9602022.}

\lref\bkk{P. Berglund, S. Katz, A. Klemm, {\it Mirror Symmetry and 
the Moduli Space for Generic
Hypersurfaces in Toric Varieties}, Nucl. Phys. B456 (1995) 153, hep-th/9506091}

\lref\candelasfont{P. Candelas, A. Font, {\it Duality Between the Webs of Heterotic and Type II Vacua}, hep-th/9603170}

\lref\dynkin{E. Perevalov, H. Skarke, {\it Enhanced Gauged Symmetry in 
Type II and F-Theory
Compactifications: Dynkin Diagrams from Polyhedra},  UTTG-15-97, 
hep-th/9704129.}

\Title{\vbox{\rightline{hep-th/9705068}\rightline{PUPT--1701 }
\rightline{}}}
{\vbox{\centerline{Exact Solutions of Exceptional Gauge Theories}
\medskip
\centerline{from Toric Geometry}}}
\bigskip
\medskip
\centerline{John H. Brodie}
\smallskip
{\it
\centerline{Department of Physics}
\centerline{Princeton University}
\centerline{Princeton, NJ 08540, USA}}
\centerline{\tt jhbrodie@princeton.edu}
\bigskip
\medskip

\vglue .3cm

\noindent
We derive four dimensional gauge theories with exceptional 
groups $F_4$, $E_8$, $E_7$, and $E_7$ with matter, by  
starting from the duality between the 
heterotic string on $K3$ and F-theory on a elliptically fibered 
Calabi-Yau 3-fold. 
This configuration is compactified to four dimensions on a torus, and  
by employing toric geometry, we compute the type IIB mirrors of the 
Calabi-Yaus of the type IIA string theory. We identify the 
Seiberg-Witten curves 
describing the gauge theories as 
ALE spaces fibered over a $P^1$ base.
\Date{4/97}

\newsec{Introduction}
Recently there has been much progress in the connection between 
gauge theories and string theory (for a review see \lerchereview). 
Although it was the gauge theory dualities
that were discovered first, it was shown later that many of them 
can be derived from string theory. 
Specifically, the quantum effects on the Coulomb branch of
$SU(2)$ and $SU(3)$ Yang-Mills theories 
can be seen directly in the duality between 
the heterotic string and F-theory, and the mirror symmetry between IIA
and IIB \refs {\kv, \selfdualstrings}. 
Initially, this correspondence between field theory and string theory
was used as a check on the string theory duality.
Later in \geomeng, the duality was dispensed with, and gauge theories 
were geometrically engineered
from type IIA and IIB theories on 
Calabi-Yaus. 
In this paper, we return and examine the duality between different 
points in the heterotic string moduli space and singularities of 
elliptically fibered Calabi-Yaus
of type IIA string theory as was mapped out in 
\refs{\bershadskyetal, \candelasfont}.
The vector moduli space of the mirror manifold of the IIB theory 
receives no quantum corrections.
By this chain of dualities, 
we examine the relation between Calabi-Yau manifolds and the
curves describing the quantum moduli space of $N=2$ four dimensional 
gauge theories based on exceptional groups.  While in \geomeng\ local
mirror symmetry was considered, we will be using the more conventional 
mirror symmetry computed using the mathematics of toric geometry
\toric. 
The advantage of this method is that, when examining theories 
where the exact solution is not known, we know what to expect of the geometry 
from the duality between type II string theory and the heterotic string.
A disadvantage is that this method
relies on always having an algebraic description of the mirror manifold. 
We will see that sometimes this requires modification of the polyhedron
associated with the Calabi-Yau 3-fold.

The enhanced gauge symmetries considered in \refs{\kv, \selfdualstrings} 
were special points of 
the moduli space of a compactification on a torus, 
and therefore involved modular functions. In contrast, 
our enhanced gauge groups 
will be coming from the ADE singular points of the $K3$ fibration 
(and outer automorphisms). There is a surprising 
mismatch between the blow-ups of the Kahler moduli corresponding to 
singularities on the IIA side and the monomials of the complex structure 
moduli that resolve the singularity on the IIB side. Moreover, it should 
be kept in mind that  
the ALE fibration are not the same as the curves of
\martinecwarner\ except for the case $A_n$, but are 
physically equivalent. This relationship has
been explored in \lerchewarner.

In section 2 we will review the results of \refs{\kv, \selfdualstrings}, 
by embedding $(12,12)$ instantons into $E_8\times E_8$ such that the 
gauge group breaks completely, and 
rederive the $SU(3)$ gauge theory 
coming from the $T^2$ compactification using toric geometry. 
In section 3, we will consider the case of $(7,17)$ instantons which 
corresponds to the group $F_4$. This is the first time that an 
exact solution to a non-simply 
laced group has been derived from a Calabi-Yau. 
As expected, the curve is an $E_6$ singularity 
whose coordinates mix with the base over which the $E_6$ is fibered
{refs\bershadskyetal, \aspinwallgross}. 
The $Z_2$ automorphism that relates $E_6$ to $F_4$ is visible in 
the Calabi-Yau. 
In section 4, we will examine 
the case with $(0,24)$ instantons which can 
give us an enhanced $E_8$ 
gauge theory. This case is of great 
interest to physicist because of it relevance to string theory 
and because it is so difficult to analyze using group theory alone.
We will see that our results match with the prediction of \selfdualstrings
of an $E_8$ singularity fibered over a ${\bf P^1}$ whose 
size corresponds to the scale of the gauge theory, $\Lambda$. 
In section 5, we will derive $E_7$ Yang-Mills gauge theory 
by embedding $(4,20)$ instantons, and in section 6, we will 
look at $(5,19)$ instantons to derive $E_7$ with a 
half a ${\bf 56}$ matter field.
In the last section, 
we will look at the case of $(9,15)$ instantons
which corresponds to $SU(3)$ gauge theory and is complicated from the 
point of view of toric geometry since the mirror naively has 
``twisted states'', and therefore the mirror manifold has no 
algebraic realization. By a somewhat ad hoc procedure, we will ``remove''
these twisted states and show that they can be expressed algebraically.

\newsec{Review of $(12,12)$ instanton embedding in $E_8\times E_8$}
\subsec{Derivation of $SU(3)$ Yang-Mills}

The heterotic string on $K3$ requires 
the embedding of 24 instantons into  
$E_8\times E_8$. In \kv, 12 instantons were embedded into
each $E_8$. The instantons are arranged to break the $E_8\times E_8$ completely. 
The number hypermultiplets is $(12-n)c_2(H) - dim(H)$ 
where $E_8 = G\times H$. $G$ 
is the group that survives the Higgsing and $H$ is the commutant.  The 
other $E_8$ factor contributes $(12+n)c_2(E_8) - dim(E_8)$. 
We add 20 for the 
moduli of the $K3$. In this case, we have $2(12\times 30 - 248)+20 = 244$ 
hypermultiplets generically.
We can compactify this six dimensional theory on a torus to get a 
four dimensional theory. The instantons ate all of the vectors of the 
$E_8\times E_8$ gauge group, but we recover two vectors from the torus in 
addition
to the dilaton and the graviphoton. 
The theory dual to this heterotic 
theory
is type IIA on $WP^4_{1,1,2,8,12}$ which has hodge numbers $h^{1,1} = 3$ and
$h^{2,1} = 243$. This type IIA theory is dual to type IIB theory on the 
mirror of  $WP^4_{1,1,2,8,12}$ which has $h^{2,1} = 3$ and
$h^{1,1} = 243$. One way to see the $h^{1,1}$ is to notice 
that $WP^4_{1,1,2,8,12}$ has
a $Z_2$ quotient singular curve $(0,0,x_3,x_4,x_5)$ 
and a $Z_4$ quotient singular point
$(0,0,0,x_4,x_5)$ which is blown-up to a Hirzebruch surface ${\bf F_2}$
\toric. The two divisors that we inserted to perform the 
blow-up plus the rescaling of the ambiant space gives $h^{1,1}=3$. 
In order to find the mirror manifold, we use toric geometry to 
construct the Newton polyhedron, 
$\Delta$, shifted by
$(-1,-1,-1,-1)$ as
\eqn\newtonpolyA{\Delta={\rm Conv}\left(\{n\in Z^5|
\sum_{i=1}^5 n_i w_i=0,\, n_i\ge -1\,\forall i\}\right),\,} where $w_i$
are the weights of the Calabi-Yau.
The corners for the 
Newton polyhedron associated with $WP^4_{1,1,2,8,12}$ are 
\eqn\cornersdeltaA{\matrix{
\nu^{(1)}=(1,-1,-1,-1)\cr
\nu^{(2)}=(-1,2,-1,-1)\cr
\nu^{(3)}=(-1,-1,11,-1)\cr
\nu^{(4)}=(-1,-1,-1,23)\cr
\nu^{(5)}=(-1,-1,-1,-1)\cr}}
There are 243 point that lie on the interior faces and edges of this polyhedron
(not including co-dimension 1 points).

This polyhedron is reflexive. The dual polyhedron is given by 
\eqn\dualpolyA{\Delta^*={\rm Conv}\left(\{m\in\Lambda^*|<n,m>\geq -1,\quad
\forall n\in\Delta\}\right)}
The dual consists of seven points plus others that correspond
to co-dimension 1 faces which do not interest us.
\eqn\cornersdeltadualA{\matrix{
\nu^{*(1)}=( 1, 0, 0, 0)\cr
\nu^{*(2)}=( 0, 1, 0, 0)\cr
\nu^{*(3)}=( 0, 0, 1, 0)\cr
\nu^{*(4)}=( 0, 0, 0, 1)\cr
\nu^{*(5)}=(-12,-6,-2,-1)\cr
\nu^{*(6)}=(-3,-2,0,0)\cr
\nu^{*(7)}=(-6,-4,-1,0)\cr}}
\ifig\Ftwo{A sketch of the dual polyhedron for $WP^4_{1,1,2,8,12}$. 
The dots represent interior points associated with divisors.}
{\epsfxsize2.0in\epsfbox{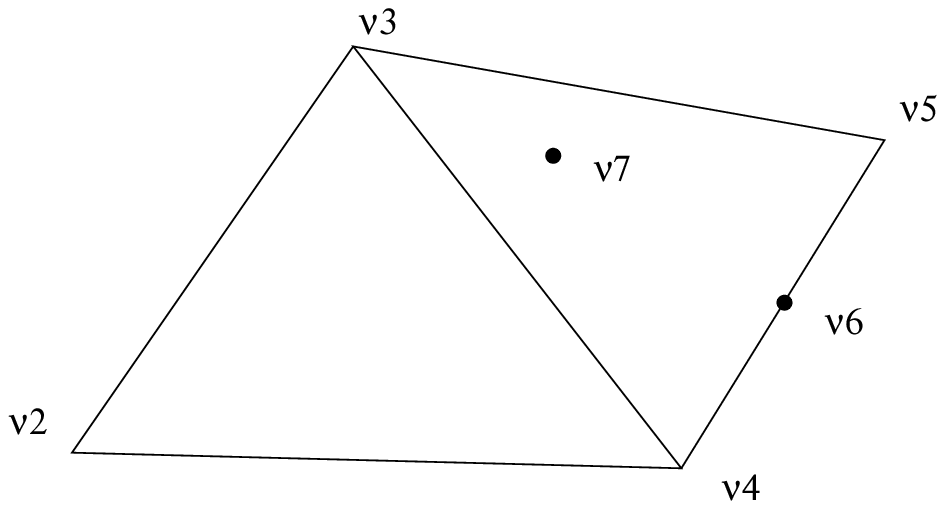}}
\ifig\Fn{${\bf F_n}$ Hirzebruch surface.}
{\epsfxsize2.0in\epsfbox{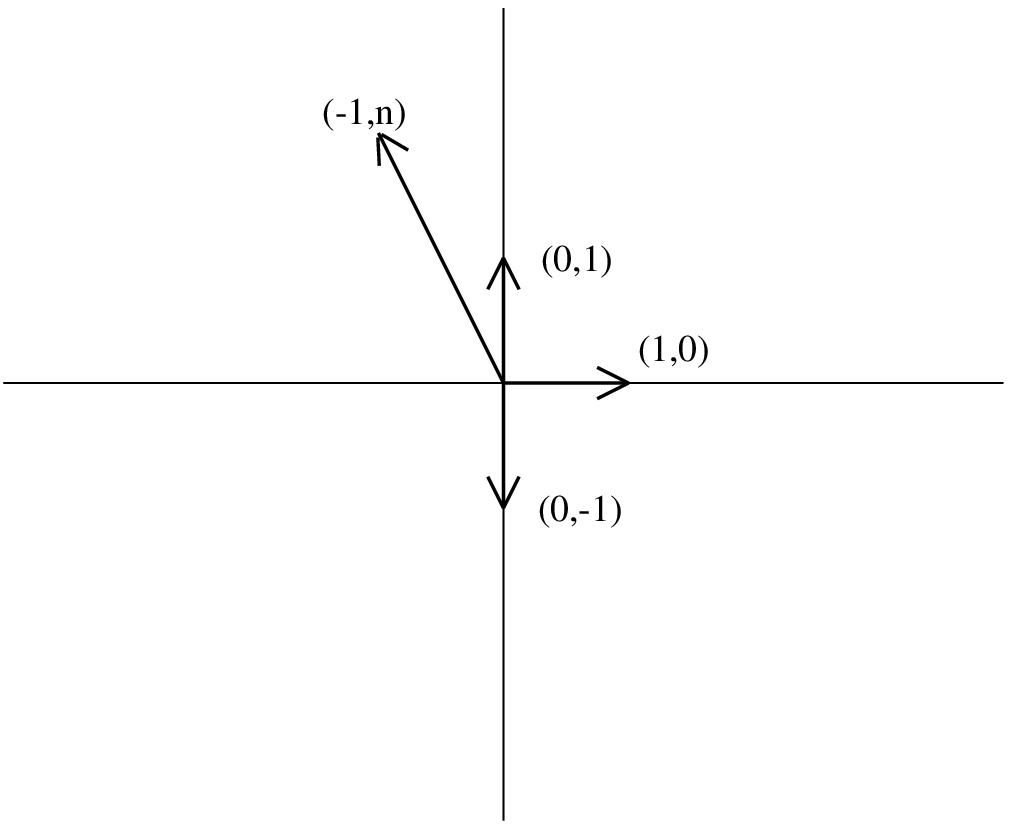}}
As is partially shown in \Ftwo, the first five points are vertices 
while the last two points correspond 
to exceptional divisors (interior points)
necessary to make the otherwise singular Calabi-Yau space smooth.
These divisors blow the base space up to be the Hirzebruch surface 
${\bf F_2}$. In fact, one can see the ${\bf F_2}$ in the two right most 
columns of \cornersdeltadualA\ (see \Fn). 
Each vertex in the dual polyhedron corresponds the power $\phi_j$ to 
which the $x_i$ is raised, $x_i^{\phi_j}$,
\eqn\bcmonomial{
\phi_j=\sum_{i\in \Lambda \cap \Delta}<(v^{(i)},1),(v^{(j)*},1)>~.}

If we compute these, we see that the mirror manifold is 
\eqn\cytwo{W = {x_1}^2 + {x_2}^3 + {x_3}^{12} + a_1{x_4}^{24} + {x_5}^{24}
+ a_0x_1x_2x_3x_4x_5 + (x_3x_4x_5)^6 + a_2(x_4x_5)^{12}}
where we have scaled the coordinates $x_i$ such that 
five of the terms have coefficient 1.
If we write this as a manifest $K3$ fibration, we take 
$\lambda = ({x_5\over x_4})^{12}$
and $x_0 = x_4x_5$. Thus, $x_4 = \sqrt{x_0\over \lambda^{1/12}}$ and
$x_5 = \sqrt{\lambda^{1/12} x_0}$. Rewriting equation \cytwo\ we find
\eqn\cytwokthreeShitf{W = {x_1}^2 + {x_2}^3 + {x_3}^{12} 
+ x_0^{12}({a_1\over \lambda} + {\lambda})
+ a_2x_0^{12} + a_0x_1x_2x_3x_0 + (x_3x_0)^6 .} 
We see that the $K3$ fiber is $WP^3_{1,1,4,6}$.
By expanding the curve about the point at which the Calabi-Yau
degenerates $W=0$ and $dW=0$, one can show 
that the $K3$ has an $A_2$ singularity 
\selfdualstrings.
Another way to see this is to redefine 
\eqn\shift{\eqalign{x_1 & = y_1 - {a_0\over 2}x_2x_3x_0\cr
	   x_2 & = y_2 + {a_0^2\over 12}x_3^2x_0^2\cr}}
\eqn\cytwokthree{W = {y_1}^2 + {y_2}^3 + {x_3}^{12} 
+ x_0^{12}({c_0\over \lambda} + {\lambda})
 - {a_0^4\over 48}y_2(x_3x_0)^4 - ({a_0^6\over 864}+1)(x_3x_0)^6
+a_2x_0^{12}.}
Going to a point where $x_0 = x_3 = 1$, we see that 
\eqn\cytwokthreeCurve{W = ({c_0\over \lambda} + {\lambda})
+ {y_1}^2 + {y_2}^3 
+ c_2y_2 + c_1.}
where $c_2 = -{a_0^4\over 48}$, $c_1 = -{a_0^6\over 864} + a_2 + 2$, 
$c_0 = a_1$.
There is an $A_2$ singularity in the coordinates $y_1$ and $y_2$ of the 
elliptic fiber.
We see that $c_0$
is the size of the $\bf P^1$ and can thus be identified with the
dilaton of the heterotic theory. $c_1$ and $c_2$ are then the 
scalars that parameterize the $SU(3)$ moduli space. One can 
check that this method is equivalent to what was done in \selfdualstrings.
The rescalings in \shift\ and for $x_3$ combined with the condition that 
\cytwokthreeCurve\ is singular is eqivalent to having the 
discriminant of the Calabi-Yau \cytwo\ vanish.

\subsec{More general instanton embeddings}
Let us consider what happens if we move $n$ instantons from one $E_8$ into 
the other $E_8$. If $n>2$, then it is possible to have 
a larger enhanced gauge group. For
$n=3$, we can have $SU(3)\times SU(3)$ in four dimensions 
where we have arranged for the second $E_8$ to be completely 
broken. For $n=4$, we can have 
$SU(3)\times SO(8)$. $n=5$ gives us $SU(3)\times F_4$. $n=6$ gives 
$SU(3)\times E_6$. $n=7$ gives 
$SU(3)\times E_7$ with a half $\bf 56$ hypermultiplet. $n=8$ gives 
$SU(3)\times E_7$. $n=9$ gives $SU(3)\times E_8\times U(1)^3$.  
$n=10$ gives $SU(3)\times E_8\times U(1)^2$. 
$n=11$ gives $SU(3)\times E_8\times U(1)$. $n=12$ gives 
$SU(3)\times E_8$ \refs{\bershadskyetal,\candelasfont}.
There are also many other gauge groups we can have if one 
considers unHiggsing the completely broken $E_8$. We will not consider 
these.

The Calabi-Yau manifolds of F-theory, dual to the 
maximally Higgsed heterotic theory with
instanton embedding $(12-n,12+n)$, are
\halign{\indent#\qquad\hfil&\hfil#\quad\hfil&\hfil#\quad\hfil&
\hfil#\quad\hfil&\hfil#\quad\hfil&\hfil#\quad\hfil&\hfil#\quad
\hfil&\hfil#\quad\hfil\cr
&$s$&$t$&$u$&$v$&$x$&$y$&$z$\cr
$\lambda$&$1$&$1$&$n$&$0$&$2n+4$&$3n+6$&$0$\cr
$\mu$&$0$&$0$&$1$&$1$&$4$&$6$&$0$\cr
$\nu$&$0$&$0$&$0$&$0$&$2$&$3$&$1$.\cr
}
\noindent
This describes a Calabi-Yau in weighted projective space 
given by $WP^4_{1,1,n,2n+4,3n+6}$ \mv.
Notice here that the second and the third rows have the same weights
as the exceptional divisors in equation \cornersdeltadualA .
This is no coincidence since we know that all of these Calabi-Yaus
are elliptical fibrations over ${\bf F_n}$ by construction. 
These exceptional divisors will appear in all of our examples, and in fact 
they correspond to the moduli of the $SU(3)$
factor that also appears in all our examples. However, this correspondence 
may not be exact on the IIB side where quantum effects are manifest.

\newsec{$(7,17)$ instantons: enhanced $F_4$}
The toric geometry corresponding to the $n=5$ theory is  
relatively simple since the 
calculation of the mirror is not complicated with the appearance of twisted
states as we will see other examples are.
The Calabi-Yau $WP^4_{1,1,5,14,21}$ is non-Fermat. 
Therefore, we expect that 
the mirror will not live in the same weighted projective space
as the original manifold. In fact, the mirror manifold 
is in weighted projective space $WP^4_{3,4,21,56,84}$.
$F_4$ is a 
non-simply laced gauge group implying that there will be some 
mixing between the coordinates of the base and the fibered $K3$ since 
$K3$ manifolds can only have simply laced $ADE$ singularities
\refs{\aspinwallgross, \bershadskyetal}.

$n=5$ has 7 instantons embedded in a $G_2$ subgroup of 
$E_8$ and 17 in the other $E_8$.
The first $E_8$ is broken to $F_4$ 
while the instantons are arranged so as to break second $E_8$ is completely. 
The number of hypermultiplets is $7\times 4 - 14 + 17\times 30 - 248+20 = 296$.
We now compactify this six dimensional theory on a torus to get a 
four dimensional theory. 
The theory dual to this
heterotic theory is, according to the table above,
type IIA on $WP^4_{1,1,5,14,21}$ which has hodge numbers $h^{1,1} = 7$ and
$h^{2,1} = 295$. This theory is dual to type IIB theory 
on the mirror manifold in $WP^4_{3,4,21,56,84}$ which has $h^{2,1} = 7$ and
$h^{1,1} = 295$ \mv. To find the mirror, 
this we can write down the corners for the 
Newton polyhedron associated with $WP^4_{1,1,5,14,21}$. 
They are 
\eqn\cornersdeltaF{\matrix{
\nu^{(1)}=(1,-1,-1,-1)\cr
\nu^{(2)}=(-1,2,-1,-1)\cr
\nu^{(3)}=(-1,-1,7,-1)\cr
\nu^{(4)}=(-1,-1,-1,41)\cr
\nu^{(5)}=(-1,-1,-1,-1)\cr
\nu^{(6)}=(-1,-1,7,1)\cr
}}
Because $WP^4_{1,1,5,14,21}$ is non-Fermat there are more than five corners.

\ifig\Ffive{A sketch of the dual polyhedron for $WP^4_{1,1,5,14,21}$. 
The dots with circles around them, represent points associated with 
the $SU(3)$ gauge theory as well as 
the divisors that blow-up the base to ${\bf F_5}$. 
The other interior points make the Dynkin diagram for 
the group $F_4$.}
{\epsfxsize2.0in\epsfbox{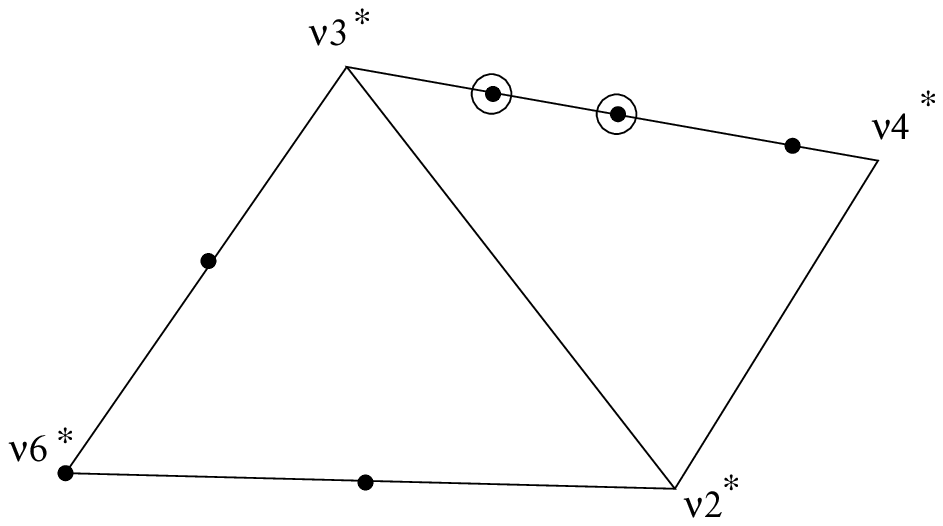}}
The dual polyhedron has eleven points.
\eqn\cornersdeltadualF{\matrix{
\nu^{*(1)}=( 1, 0, 0, 0)\cr
\nu^{*(2)}=( 0, 1, 0, 0)\cr
\nu^{*(3)}=( 0, 0, 1, 0)\cr
\nu^{*(4)}=( 0, 0, 0, 1)\cr
\nu^{*(5)}=(-21,-14,-5,-1)\cr
\nu^{*(6)}=(-12,-8,-3,0)\cr
\nu^{*(7)}=(-9,-6,-2,0)\cr
\nu^{*(8)}=(-8,-5,-2,0)\cr
\nu^{*(9)}=(-6,-4,-1,0)\cr
\nu^{*(10)}=(-3,-2,0,0)\cr
\nu^{*(11)}=(-4,-2,-1,0)\cr
}}
where the first six points are vertices while the 
last five are point lie the interior of the polyhedron 
as shown in \Ffive. 
As was also pointed out in \dynkin\ the points on the polyhedron are 
in the configuration of an $F_4$ Dynkin diagram.

Using \bcmonomial\ to compute 
the corresponding monomials, we see that the mirror manifold is 
\eqn\cyFfour{\eqalign{W = & {x_1}^2 + {x_2}^3 + {x_3}^{8} 
+ x_3^2{x_4}^{42} + {x_5}^{42} 
+ a_0x_1x_2x_3x_4x_5 + a_1(x_3x_4x_5)^6 + a_2x_3^4(x_4x_5)^{12}\cr
+ & a_3x_2^2(x_4x_5)^{8} 
+ a_4x_3^2(x_4x_5)^{18} + a_5(x_4x_5)^{24} + a_6x_2(x_4x_5)^{16}.\cr}}
Notice that this equation is invariant under $x_3 \rightarrow -x_3$
and $x_1 \rightarrow -x_1$.
Finding the discriminant and seeing where it vanishes 
is very complicated for $F_4$, but we can try another tact.
By introducing a redefinition of coordinates that preserves the 
weights of the Calabi-Yau, 
\eqn\zeroF{\eqalign{x_2 & = {a_0^2\over 12}x_3^2(x_4x_5)^2 
		   - {a_3\over 3}(x_4x_5)^{8} + y_2\cr
		   x_1 & = -{a_0\over 2}x_2x_3(x_4x_5) + y_1\cr}}
the Calabi-Yau takes the form 
\eqn\cyFfourmod{\eqalign{W = & {y_1}^2 + {y_2}^3 + {y_3}^{8} 
+ x_3^2{x_4}^{42} + {x_5}^{42} \cr
+ & c_1y_3^4y_2(x_4x_5)^4 + c_2y_3^6(x_4x_5)^6 
+ c_0y_3^4(x_4x_5)^{12} 
+ c_3y_3^2y_2(x_4x_5)^{10} \cr
+ & c_4y_3^2(x_4x_5)^{18} 
+ c_5(x_4x_5)^{24} + c_6y_2(x_4x_5)^{16}.\cr}}
where
\eqn\cs{\eqalign{c_0 & = a_2 + {a_0^4a_3\over 72}\cr
		 c_1 & = -{a_0^4\over 48}\cr
		 c_2 & = -{a_0^6\over 864} + a_1\cr
		 c_3 & =  {a_0^2a_3\over 6}\cr
                 c_4 & = -{a_0^2a_3^2\over 18} 
			+ a_4 + {a_6a_0^2\over 12}\cr
	  	 c_5 & =  a_5+{2a_3^3\over 27} - {a_6a_3\over 3}\cr
		 c_6 & = -{a_3^2\over 3} + a_6.\cr}}
Now we rescale $x_4 = {y_4\over c_0^{1/12}}$ 
and define $\lambda = ({x_5\over x_4})^{24}$ 
and $x_0 = y_4x_5$.
Going to the point where $x_0 = 1$, we see that we have an $E_6$ singularity 
fibered over a ${\bf P^1}$ base just as was predicted 
in \selfdualstrings. 
\eqn\cyFfourcurve{\eqalign
{W = &   {x_3^2\over c_0^{7/2}\lambda} + \lambda + {y_3}^{8}
+  {c_1\over c_0^{1/3}}y_3^4y_2 + {c_2\over c_0^{1/2}}y_3^6 
+  y_3^4 +  {y_2}^3 + y_1^2 \cr
+  & {c_3\over c_0^{5/6}}y_3^2y_2 + {c_4\over c_0^{3/2}}y_3^2  
+ {c_5\over c_0^2} + {c_6\over c_0^{4/3}}y_2.\cr}}
Notice that the coordinates of the 
fiber mix with the base. Notice also that there is an additional 
${\bf Z_2}$ symmetry inhereted from the Calabi-Yau 
that is not present in the usual $E_6$ curve $y_3 \rightarrow -y_3$ and 
$y_1 \rightarrow -y_1$ that reduces the number of monomials 
from six to four.
This is the ${\bf Z_2}$ automorphism that takes $E_6$ to $F_4$
\refs{\aspinwallgross, \bershadskyetal}.

The terms $y_3^8$, $y_3^6$, and $y_3^4y_2$ are all irrelevant operators
near the $E_6$ singularity. 
We can identify ${1\over c_0^{7/2}}$ with the size of the 
${\bf P^1}$ and therefore the scale of the $F_4$ gauge theory.
We identify ${c_i\over c_0^{\chi_i}}$ 
for $i=3\ldots 6$ with the four Casimirs of the 
$F_4$ gauge group (where $\chi_i$ is given in equation
\cyFfourcurve).
What we have essentially done is expand 
the coordinates about the point at which the Calabi-Yau degenerates
($W = 0$, $dW=0$) 
and we have found that there is an $F_4$ singularity at that point. 

It is somewhat surprising that we had to redefine our variables 
(equation \zeroF ) to see the $F_4$ singularity since the 
terms with coefficients $a_i$ are a map of the blow ups of the 
$F_4$ singularity on the IIA side. Apparently there are 
quantum effects on the IIB side that force us to introduce 
a redefinition of variables (equation \zeroF ) such that 
we can see that there indeed is an $F_4$ singularity.

It would be interesting to show that the 2-cycles associated with 
the $F_4$ singularity can be traced back to the 3-cycles of this 
Calabi-Yau.

One might also ask where is the $SU(3)$ of the $SU(3)\times F_4$. 
The moduli must be related to the two moduli ${c_1\over c_0^{1/3}}$ and 
${c_2\over c_0^{1/2}}$.
The $A_2$ singularity must also be infinitely far away from the 
$E_8$ singularity on the Calabi-Yau.

Also note that equation \cyFfourcurve\ is not a Riemann surface; 
however, it is 
presumably physically equivalent to such a construction base
on integrable systems and Prym varieties as discussed in 
\refs{\martinecwarner \lerchewarner}. 

\newsec{$(0,24)$ instantons: enhanced $E_8$}
Let's consider what happens if $n=12$, this is the case where 
there are zero instantons embedded in one $E_8$ and 24 instantons 
in the other
 such that the group breaks completely.
$E_8$ is a simply-laced gauge group
so we expect the singularity in the $K3$ fiber not to mix with the
base ${\bf P^1}$. 
On the other hand, $E_8$ is a difficult case to consider 
since the complete list of 
classical invariants is unknown \refs{\giddings, \cvitanovic}.

The first $E_8$ is unbroken while the second $E_8$ is completely broken. 
The 
number of hypermultiplets is $24\times 30 - 248+20 = 492$.
Compactification of this six dimensional theory on a torus yields a 
four dimensional theory. 
The theory dual to this
heterotic theory is, according to the table above,
type IIA on $WP^4_{1,1,12,28,42}$ with hodge numbers $h^{1,1} = 11$ and
$h^{2,1} = 491$. This type IIA theory is dual to type IIB theory on the 
mirror of  $WP^4_{1,1,12,28,42}$. To find the mirror, 
this we can write down the corners for the 
Newton polyhedron associated with $WP^4_{1,1,12,28,42}$. They are 
\eqn\cornersdeltaEeight{\matrix{
\nu^{(1)}=(1,-1,-1,-1)\cr
\nu^{(2)}=(-1,2,-1,-1)\cr
\nu^{(3)}=(-1,-1,6,-1)\cr
\nu^{(4)}=(-1,-1,-1,83)\cr
\nu^{(5)}=(-1,-1,-1,-1)\cr}}

\ifig\Ftwelve{A sketch of the dual polyhedron for $WP^4_{1,1,12,28,42}$. 
The dots with circles are associated with the ${\bf F_{12}}$ blow-up 
and the $SU(3)$ gauge theory. 
The other dots are interior points and make the Dynkin diagram of $E_8$.}
{\epsfxsize2.0in\epsfbox{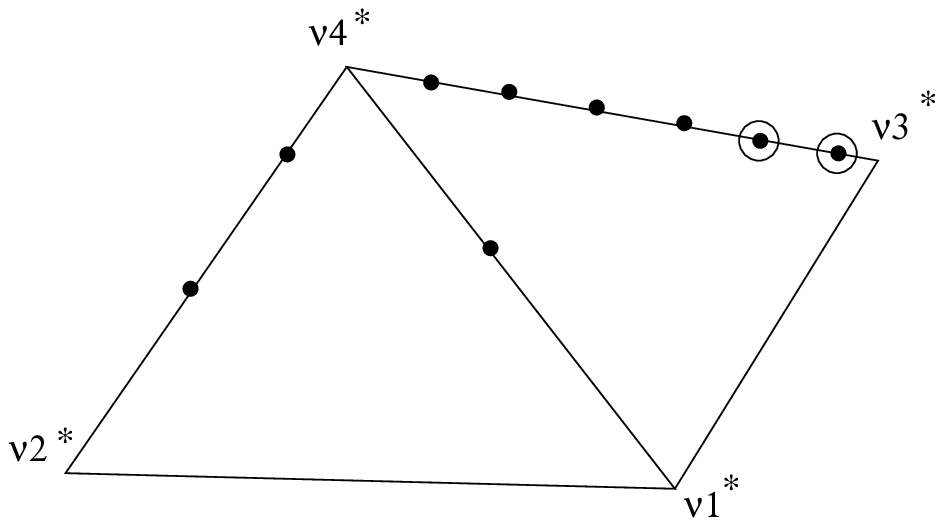}}
As shown in \Ftwelve, the dual polyhedron has fifteen points.
\eqn\cornersdeltadual{\matrix{
\nu^{*(1)}=( 1, 0, 0, 0)\cr
\nu^{*(2)}=( 0, 1, 0, 0)\cr
\nu^{*(3)}=( 0, 0, 1, 0)\cr
\nu^{*(4)}=( 0, 0, 0, 1)\cr
\nu^{*(5)}=(-42,-28,-12,-1)\cr
\nu^{*(6)}=(-3,-2,0,0)\cr
\nu^{*(7)}=(-6,-4,-1,0)\cr
\nu^{*(8)}=(-7,-4,-2,0)\cr
\nu^{*(9)}=(-9,-6,-2,0)\cr
\nu^{*(10)}=(-10,-7,-3,0)\cr
\nu^{*(11)}=(-12,-8,-3,0)\cr
\nu^{*(12)}=(-14,-9,-4,0)\cr
\nu^{*(13)}=(-15,-10,-4,0)\cr
\nu^{*(14)}=(-18,-12,-5,0)\cr
\nu^{*(15)}=(-21,-14,-6,0)\cr
}}
where the last ten points correspond to exceptional divisors
necessary to make the otherwise singular space smooth
(there are
also other points that correspond to faces of co-dimension 1 which 
do not concern us since they don't make the space singular). 

Using \bcmonomial\ to compute the monomials, 
we see that the mirror manifold is 
\eqn\cyEeight{\eqalign{W = & {x_1}^2 + {x_2}^3 + {x_3}^{7} 
+ {x_4}^{84} + {x_5}^{84} 
+ a_0x_1x_2x_3x_4x_5 + a_1(x_3x_4x_5)^6 \cr
+ & a_2x_3^5(x_4x_5)^{12} + a_3x_2^2(x_4x_5)^{14} 
+ a_4x_3^4(x_4x_5)^{18}+ a_5x_3^3(x_4x_5)^{24} \cr
+ & a_6x_3^2(x_4x_5)^{30} +  a_7x_3(x_4x_5)^{36} + a_8(x_4x_5)^{42}
+ a_9x_2(x_4x_5)^{28} + a_{10}x_1(x_4x_5)^{21}.\cr}}
By introducing a definition of coordinates that preserves the 
weights of the Calabi-Yau, 
\eqn\zeroE{\eqalign{x_3 & = b_1(x_4x_5)^6 + b_2y_3 \cr
		   x_2 & = b_3x_3^2(x_4x_5)^2 + b_4(x_4x_5)^{14} + b_5y_2\cr
		   x_1 & = b_6x_2x_3(x_4x_5) + b_7(x_4x_5)^{21} + b_8y_1}}
and completing the square, the cube, and the septet
the Calabi-Yau takes the form 
\eqn\cyEeight{\eqalign{W = & {y_1}^2 + {y_2}^3 + {y_3}^{7} 
+ c_0{\tilde x_4}^{84} + {x_5}^{84} \cr
+ & c_1y_3^4y_2(\tilde x_4x_5)^4 + c_2y_3^6(\tilde x_4x_5)^6 
+ y_3^5(\tilde x_4x_5)^{12} 
+ c_3y_3y_2(\tilde x_4x_5)^{22} \cr
+ & c_4y_3^3y_2(\tilde x_4x_5)^{16}+ c_5y_3^3(\tilde x_4x_5)^{24} 
+ c_6y_3^2(\tilde x_4x_5)^{30} +  c_7y_3(\tilde x_4x_5)^{36} \cr
+ & c_8(\tilde x_4x_5)^{42} + c_9y_2(\tilde x_4x_5)^{28} 
+ c_{10}y_3^2y_2(\tilde x_4x_5)^{16}.\cr}}
where we have rescaled $x_4$ such that the term $y_3^5$ has coefficient 1.
Now we define $\lambda = ({x_5\over \tilde x_4})^{42}$ 
and $x_0 = \tilde x_4x_5$.
Going to the point where $x_0 = 1$, 
we see that we have an $E_8$ singularity 
fibered over a ${\bf P^1}$ base.
\eqn\cyEeight{\eqalign{W = &  
 {c_0\over \lambda} + \lambda + {y_3}^{7}
+  c_1y_3^4y_2 + c_2y_3^6 
+  y_3^5 +  y_1^2 + {y_2}^3 + c_3y_3y_2 \cr
+  & c_4y_3^3y_2 + c_5y_3^3 + c_6y_3^2 +  c_7y_3 + c_8
+ c_9y_2 + c_{10}y_3^2y_2.\cr}}
The terms $y_3^7$, $y_3^6$, and $y_3^4y_2$ are all irrelevant operators
near the $E_8$ singularity. The parameters $c_i$ depend on $a_j$.
We can identify $c_0$ with the size of the ${\bf P^1}$ and 
therefore the scale of the $E_8$ gauge theory $\Lambda$.
$c_i$ for $i=3\ldots 10$ we identify with the eight Casimirs of the 
$E_8$ gauge group.

We are not aware that 
a Riemann surface corresponding to the moduli space of 
an $N=2$ $E_8$ gauge theory exists in the physics literature at this 
time although the recepe for its construction was given in 
\martinecwarner.

\newsec{$(4,20)$ instantons: enhanced $E_7$}
$n=8$ has 4 instantons embedded in an $SU(2)$ subgroup of one $E_8$ 
and 20 in the other $E_8$.
The first $E_8$ is broken to $E_7$ 
while we arrange the instantons to completely break the second $E_8$. The 
number of hypermultiplets is $4\times 2 - 3 + 20\times 30 - 248+20 = 377$.
We now compactify this six dimensional theory on a torus to get a 
four dimensional theory. 
The theory dual to this
heterotic theory is, according to the table above,
type IIA on $WP^4_{1,1,8,20,30}$ which has hodge numbers $h^{1,1} = 10$ and
$h^{2,1} = 376$. This theory is dual to type IIB theory 
on the mirror manifold in $WP^4_{1,1,12,16,30}$. To find the mirror, 
this we can write down the corners for the 
Newton polyhedron associated with $WP^4_{1,1,8,20,30}$. 
They are 
\eqn\cornersdeltaE{\matrix{
\nu^{(1)}=(1,-1,-1,-1)\cr
\nu^{(2)}=(-1,2,-1,-1)\cr
\nu^{(3)}=(-1,-1,6,-1)\cr
\nu^{(4)}=(-1,-1,6, 3)\cr
\nu^{(5)}=(-1, 0,4,-1)\cr
\nu^{(6)}=(-1,-1,-1,59)\cr
\nu^{(7)}=(-1,-1,-1,-1)\cr
}}

\ifig\Feight{A sketch of the dual polyhedron for $WP^4_{1,1,8,20,30}$. 
The dots represent interior points associated with monomials of $E_7$.}
{\epsfxsize2.0in\epsfbox{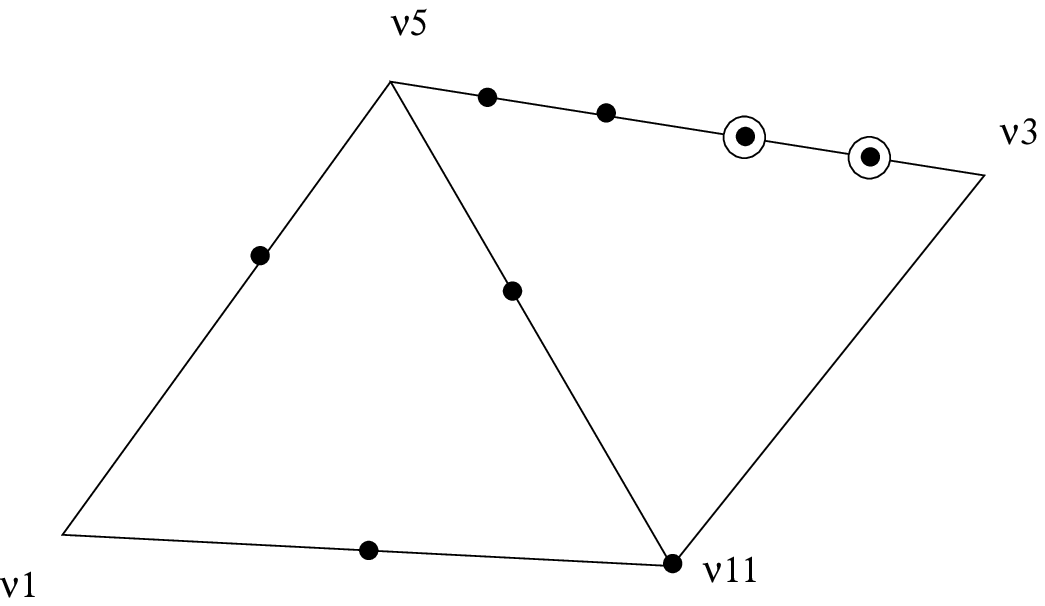}}
As shown in \Feight, the dual polyhedron consists of the following points.
\eqn\cornersdeltadualE{\matrix{
\nu^{*(1)}=( 1, 0, 0, 0)\cr
\nu^{*(2)}=( 0, 1, 0, 0)\cr
\nu^{*(3)}=( 0, 0, 1, 0)\cr
\nu^{*(4)}=( 0, 0, 0, 1)\cr
\nu^{*(5)}=(-30,-20,-8,-1)\cr
\nu^{*(6)}=(-15,-10,-4,0)\cr
\nu^{*(7)}=(-12,-8,-3,0)\cr
\nu^{*(8)}=(-11,-7,-3,0)\cr
\nu^{*(9)}=(-9,-6,-2,0)\cr
\nu^{*(10)}=(-7,-5,-2,0)\cr
\nu^{*(11)}=(-7,-4,-2,0)\cr
\nu^{*(12)}=(-6,-4,-1,0)\cr
\nu^{*(13)}=(-3,-2,-1,0)\cr
\nu^{*(14)}=(-3,-2,0,0)\cr
}}

Using \bcmonomial\ to compute 
the corresponding monomials, we see that the mirror manifold is 
\eqn\cyEseven{\eqalign{W = & {x_1}^2 + {x_2}^3x_3 + {x_3}^{5} 
+ {x_4}^{60} + {x_5}^{60} 
+ a_0x_1x_2x_3x_4x_5 + a_1x_3^4(x_4x_5)^6 \cr
+ & a_2x_1x_2(x_4x_5)^{7} + a_3x_2^2(x_4x_5)^{14} \cr
+ & a_4x_3^3(x_4x_5)^{12} + a_5(x_4x_5)^{30} + a_6x_2(x_4x_5)^{22}\cr
+ & a_7x_3(x_4x_5)^{24} + a_8x_3^2(x_4x_5)^{18} + a_9x_1(x_4x_5)^{15}
.\cr}}
This Calabi-Yau is embedded in the space $WP^4_{1,1,12,16,30}$ 
moded out by the discrete group of the mirror.
By introducing a definition of coordinates that preserves the 
weights of the Calabi-Yau, 
\eqn\zeroEseven{\eqalign{x_3 & = b_1(x_4x_5)^6 + b_2y_3 \cr
		   x_2 & = b_3x_3(x_4x_5)^2 + b_4(x_4x_5)^{8} + b_5y_2\cr
		   x_1 & = b_6x_2x_3(x_4x_5) + b_7x_2(x_4x_5)^7 + b_8y_1}}
completing the square, the cube, and the quartic, 
the Calabi-Yau takes the form 
\eqn\cyEsevenMod{\eqalign{W = & {y_1}^2 + {y_2}^3y_3 + {y_3}^{5} 
+ c_0{\tilde x_4}^{60} + {x_5}^{60} 
+ c_1y_3^2y_2^2(\tilde x_4x_5)^2 + c_2y_3^4(\tilde x_4x_5)^6 \cr
+ & y_3^3(\tilde x_4x_5)^{12} 
+ c_3y_3^2y_2(\tilde x_4x_5)^{10} 
+ c_4y_3y_2(\tilde x_4x_5)^{16}
+ c_5y_3^2(\tilde x_4x_5)^{18} 
+ c_6y_2(\tilde x_4x_5)^{22}\cr
+ & c_7y_3(\tilde x_4x_5)^{24}
+ c_8y_2^2(\tilde x_4x_5)^{28}
+ c_9(\tilde x_4x_5)^{30}
.\cr}}
The modulus in front of $x_4^{60}$ was 
swapped for the modulus in front of the $x_3^3$ term
by rescaling $x_4$.
Now we define $\lambda = ({x_5\over \tilde x_4})^{60}$ 
and $x_0 = \tilde x_4x_5$.
Going to the point where $x_0 = 1$, we see that we have an $E_7$ singularity 
fibered over a ${\bf P^1}$ base. 
\eqn\cyEsevenCurve{\eqalign{
W = & {c_0\over \lambda} + \lambda 
+ c_1y_3^2y_2^2 + c_2y_3^4 + {y_3}^{5}\cr
+ & {y_1}^2 + {y_2}^3y_3  
+ y_3^3 + c_3y_3^2y_2
+ c_4y_3y_2 + c_5y_3^2
+ c_6y_2 + c_7y_3 + c_8y_2^2 + c_9
}}
The terms $y_3^5$, $y_3^4$, and $y_3^2y_2^2$ are all irrelevant operators
near the $E_6$ singularity. The parameters $c_i$ depend on $a_j$.
We can identify $c_0$ with the size of the ${\bf P^1}$ and 
therefore the scale of the $E_7$ gauge theory.
$c_i$ for $i=3\ldots 9$ we identify with the seven Casimirs of the 
$E_7$ gauge group.

\newsec{$(5,19)$ instantons: enhanced $E_7$ with matter}
$n=7$ has 5 instantons embedded an $SU(2)$ subgroup of one $E_8$ 
and 19 in the other $E_8$.
The first $E_8$ is broken to $E_7$ with ${1\over 2}{\bf 56}$ 
while the second $E_8$ is completely broken. The 
number of hypermultiplets is 
$4\times 2 - 3 + 20\times 30 - 248+20-28 = 349$.
The theory dual to this
heterotic theory is
type IIA on $WP^4_{1,1,7,18,27}$ which has hodge numbers $h^{1,1} = 10$ and
$h^{2,1} = 348$. To find the mirror, 
this we can write down the corners for the 
Newton polyhedron associated with $WP^4_{1,1,7,18,27}$. 
They are 
\eqn\cornersdeltaEseven{\matrix{
\nu^{(1)}=(1,-1,-1,-1)\cr
\nu^{(2)}=(-1,2,-1,-1)\cr
\nu^{(3)}=(-1,-1,6,-1)\cr
\nu^{(4)}=(-1,-1,6, 4)\cr
\nu^{(5)}=(-1, 0,4,-1)\cr
\nu^{(6)}=(-1, 0,4, 0)\cr
\nu^{(7)}=(-1,-1,-1,59)\cr
\nu^{(8)}=(-1,-1,-1,-1)\cr
}}

The dual polyhedron has fourteen points.
\eqn\cornersdeltadualEseven{\matrix{
\nu^{*(1)}=( 1, 0, 0, 0)\cr
\nu^{*(2)}=( 0, 1, 0, 0)\cr
\nu^{*(3)}=( 0, 0, 1, 0)\cr
\nu^{*(4)}=( 0, 0, 0, 1)\cr
\nu^{*(5)}=(-27,-18,-7,-1)\cr
\nu^{*(6)}=(-15,-10,-4,0)\cr
\nu^{*(7)}=(-12,-8,-3,0)\cr
\nu^{*(8)}=(-11,-7,-3,0)\cr
\nu^{*(9)}=(-9,-6,-2,0)\cr
\nu^{*(10)}=(-7,-5,-2,0)\cr
\nu^{*(11)}=(-7,-4,-2,0)\cr
\nu^{*(12)}=(-6,-4,-1,0)\cr
\nu^{*(13)}=(-3,-2,-1,0)\cr
\nu^{*(14)}=(-3,-2,0,0)\cr
}}

If we use \bcmonomial\ to compute 
the corresponding monomials, we see that the mirror manifold is 
\eqn\cyMatter{\eqalign{W = & {x_1}^2 + {x_2}^3x_3 + {x_3}^{5} 
+ x_3{x_4}^{54} + {x_5}^{54} 
+ a_0x_1x_2x_3x_4x_5 + a_1x_3^4(x_4x_5)^6 \cr
+ & a_2x_1x_2(x_4x_5)^{7} + a_3x_2^2(x_4x_5)^{14} \cr
+ & a_4x_3^3(x_4x_5)^{12} + a_5(x_4x_5)^{30} + a_6x_2(x_4x_5)^{22}\cr
+ & a_7x_3(x_4x_5)^{24} + a_8x_3^2(x_4x_5)^{18} + a_9x_1(x_4x_5)^{15}
.\cr}}
This polynomial is just the same as \cyEseven\ except for the 
term $x_3{x_4}^{54}$. The redefinitions are the same, and we have
\eqn\cyEmattercurve{\eqalign{
W = & {c_0x_3\over \lambda} + \lambda + c_1y_3^2y_2^2 + c_2y_3^4 
+ {y_3}^{5} + {y_1}^2 + {y_2}^3y_3  + y_3^3 \cr
+ & c_3y_3^2y_2
+ c_4y_3y_2 + c_5y_3^2 
+ c_6y_2 + c_7y_3 + c_8y_2^2 + c_9
}}
This curve is just the same as the curve for $E_7$ Yang-Mills except 
that the fiber mixes with the base. This is just what one expects. 
Presumably for $N_f$ ${1\over 2}{\bf 56}$ fundamentals 
the mixing of the ALE space and the ${\bf P^1}$ would be 
of the form ${x_3^{N_f}\over \lambda}$.

\newsec{$(9,15)$ instantons leaving $SU(3)$}
Let's consider what happens if $n=3$, that is the case where 
there are 9 instantons embedded in one $E_8$ and 15 in the other $E_8$.
The $E_6$ subgroup of the first $E_8$ is broken leaving $SU(3)$ while 
we arrange for the instantons to break the second $E_8$ is completely. 
The number of hypermultiplets is 
$9\times 12 - 78 + 15\times 30 - 248+20 = 252$.
We now compactify this six dimensional theory on a torus to get a 
four dimensional theory. 
The theory dual to this
heterotic theory is, according to the table above,
type IIA on $WP^4_{1,1,3,10,15}$ which has hodge numbers $h^{1,1} = 5$ and
$h^{2,1} = 251$. This theory is dual to type IIB theory on the 
mirror of  $WP^4_{1,1,3,10,15}$. To find the mirror, 
this we can write down the corners for the 
Newton polyhedron associated with $WP^4_{1,1,3,10,15}$. They are 
\eqn\cornersdeltaAA{\matrix{
\nu^{(1)}=(1,-1,-1,-1)\cr
\nu^{(2)}=(-1,2,-1,-1)\cr
\nu^{(3)}=(-1,-1,9,-1)\cr
\nu^{(4)}=(-1,-1,-1,29)\cr
\nu^{(5)}=(-1,-1,-1,-1)\cr}}

The dual polyhedron has eight points.
\eqn\cornersdeltadualAA{\matrix{
\nu^{*(1)}=( 1, 0, 0, 0)\cr
\nu^{*(2)}=( 0, 1, 0, 0)\cr
\nu^{*(3)}=( 0, 0, 1, 0)\cr
\nu^{*(4)}=( 0, 0, 0, 1)\cr
\nu^{*(5)}=(-15,-10,-3,-1)\cr
\nu^{*(6)}=(-3,-2,0,0)\cr
\nu^{*(7)}=(-6,-4,-1,0)\cr
\nu^{*(8)}=(-5,-3,-1,0)\cr}}

If we use \bcmonomial\ to compute the corresponding monomials, we see that the mirror manifold is 
\eqn\cythree{\eqalign{W & = {x_1}^2 + {x_2}^3 + {x_3}^{10} + {x_4}^{30} + {x_5}^{30}
+ a_0x_1x_2x_3x_4x_5 + a_1(x_3x_4x_5)^6 \cr
+ & a_2x_3^2(x_4x_5)^{12} 
+ a_3x_2(x_4x_5)^{10}.}}

\subsec{twisted states}
However, there is a problem. We said that the mirror has $h^{2,1}=5$,
but we only see four complex structure moduli in equation 
\cythree. This is because one 
of the states is a so called ``twisted state''. One can see this 
in the toric geometry, by noting that $\nu^{*(8)}$ lies on the
face $(\nu^{*(2)},\nu^{*(4)},\nu^{*(5)})$. The dual of this face is
$(\nu^{(1)},\nu^{(3)})$ which contains the point 
$\tilde \nu^{(3)} = {\nu^{(1)}+\nu^{(3)}\over 2} = (0,-1,4,-1)$. 
An algebraic description of the five complex structure moduli 
can sometimes be found by replacing
$\nu^{(3)}$ by $\tilde \nu^{(3)}$. Taking
as the corners of the polyhedron
\eqn\cornersdeltaAAt{\matrix{
\nu^{(1)}=(1,-1,-1,-1)\cr
\nu^{(2)}=(-1,2,-1,-1)\cr
\nu^{(3)}=(0,-1,4,-1)\cr
\nu^{(4)}=(-1,-1,-1,29)\cr
\nu^{(5)}=(-1,-1,-1,-1)\cr}}

This dual polyhedron has {\it nine} points.
\eqn\cornersdeltadualAAt{\matrix{
\nu^{*(1)}=( 1, 0, 0, 0)\cr
\nu^{*(2)}=( 0, 1, 0, 0)\cr
\nu^{*(3)}=( 0, 0, 1, 0)\cr
\nu^{*(4)}=( 0, 0, 0, 1)\cr
\nu^{*(5)}=(-15,-10,-3,-1)\cr
\nu^{*(6)}=(-3,-2,0,0)\cr
\nu^{*(7)}=(-6,-4,-1,0)\cr
\nu^{*(8)}=(-5,-3,-1,0)\cr
\nu^{*(9)}=(-4,-3,-1,0)\cr}}
Including the point at the origin $(0,0,0,0)$ this Calabi-Yau has the 
right 
Hodge numbers to be the mirror manifold.

If we use \bcmonomial\ to compute the corresponding monomial, 
we see that the mirror manifold is 
\eqn\cythree{\eqalign{W & 
= {x_1}^2x_3 + {x_2}^3 + {x_3}^{5} + {x_4}^{30} + {x_5}^{30} 
+ a_0x_1x_2x_3x_4x_5 + a_1x_3^3(x_4x_5)^6 \cr 
+ & a_2x_3(x_4x_5)^{12} + a_3x_2(x_4x_5)^{10} + a_4x_1(x_4x_5)^9.\cr}}
Note that this weighted projective space 
$WP^4_{12,10,6,1,1}$ is different from the one we started with.
Redefining the coordinates leads to an $A_2$ singularity.

\newsec{Conclusions}
We have seen how the curves describing the Coulomb branch of 
$N=2$ four dimensional Yang-Mills theories are described by 
ALE fibrations. This realizes the proposal of 
\selfdualstrings\ for the groups $F_4$, $E_8$, $E_7$, and $E_7$ with matter. 
Calabi-Yaus prove to be a powerful tool for studying gauge theories.
Surprisingly,
we have seen that the complex structure moduli of the IIB theory 
associated with the blow ups of the Kahler singularities of the 
mirror IIA theory do not match up with the moduli of the ADE type
singularity. We have seen that combinations of the the blow ups of 
the Kahler singularities can be arranged such that complex structure
moduli of the IIB theories do indeed give up the ALE fibrations that
we expect. This is a non-trivial check of the string theory 
realization of gauge symmetry.

In the broader context, it is in some ways surprising that Calabi-Yaus 
have anything at all to do with 
gauge theories. Calabi-Yaus being compact space-time only make sense in 
the context of general relativity. Non-Abelian gauge fields exist independent
of gravity. Nevertheless, we see that by looking at the points at which the 
Calabi-Yaus degenerate, we see enhanced gauge fields. We are of course
looking at the limit in which $\alpha '$ is going to zero. 
However, since gauge theory 
dualities are seen more naturally as consequences of string theory, we might 
ask why a theory that requires the existence of quantum gravity should 
also require gauge theory dualities. 

With recent success made with D-branes on 
a flat background in the 
context of $N=1$ dualities one might wonder what role 
Calabi-Yaus will play beyond $N=2$. In some sense, it is not surprising 
that Calabi-Yaus were related to the elliptic curves that describe the
Coulomb branches of $N=2$ theories and not the Higgs branches of $N=1$ 
theories. Will there be a pure geometrical description of Seiberg's 
duality? 
Of course, Higgs branches aren't the only interesting phases of 
$N=1$ theories. Certainly, Calabi-Yau will have something to say 
about these other phases.

\newsec{Acknowledgements}
I would like to thank P. Aspinwall, B. Greene, K. Intriligator, S. Kachru,
D. Kaplan, A. Lorentz, P. Mayer, D. Morrison, and C. Vafa
for useful discussions.

\listrefs

\end